\documentclass[twocolumn,prb,english,showpacs,preprintnumbers,amsmath,amssymb]{revtex4}

\usepackage{graphicx}% Include figure files
\usepackage{dcolumn}% Align table columns on decimal point
\usepackage{bm}% bold math

\usepackage[latin1]{inputenc}
\usepackage{amsmath}
\usepackage{babel}
\usepackage{graphics}
\usepackage{amssymb}
\newcommand{\MnGa}{Mn$_{\rm Ga}$\ }
\newcommand{\JMn}{J_n^{\rm Mn-Mn}}
\newcommand{\up}{\uparrow}
\newcommand{\dn}{\downarrow}
\newcommand{\sv}{{\bf S}}

\begin{document}

%\draft

\title{Disorder and the effective Mn-Mn exchange interaction in\\ Ga$_{1-x}$Mn$_x$As diluted magnetic semiconductors}

\author{Ant\^onio J. R. da Silva,$^1$ A. Fazzio,$^1$ Raimundo R. dos Santos,$^2$ and Luiz E. Oliveira$^3$}

\affiliation{$^1$Instituto de F\'\i sica, Universidade de S\~ao Paulo, CP
66318, 05315-970, S\~ao Paulo SP, Brazil\\
$^2$Instituto de F\'\i sica, Universidade Federal do
Rio de Janeiro, CP 68528, 21945-970 Rio de Janeiro RJ, Brazil\\
$^3$Instituto de F\'{\i}sica, Unicamp, CP 6165,
13083-970 Campinas SP, Brazil}

\begin{abstract}
We perform a theoretical study, using {\it ab initio} total energy
density-functional calculations, of the effects of disorder on the
$Mn-Mn$ exchange interactions for $Ga_{1-x}Mn_xAs$ diluted
magnetic semiconductors. For a 128 atoms supercell, we consider a
variety of configurations with 2, 3 and 4 Mn atoms, which
correspond to concentrations of 3.1\%, 4.7\%, and 6.3\%,
respectively. In this way, the disorder is intrinsically
considered in the calculations. Using a Heisenberg Hamiltonian to
map the magnetic excitations, and {\it ab initio} total energy
calculations, we obtain the effective $\JMn$, from first ($n=1$)
all the way up to sixth ($n=6$) neighbors. Calculated results show
a clear dependence in the magnitudes of the $\JMn$ with the Mn
concentration $x$. Also, configurational disorder and/or
clustering effects lead to large dispersions in the Mn-Mn exchange
interactions, in the case of fixed Mn concentration. Moreover,
theoretical results for the ground-state total energies for
several configurations indicate the importance of a proper
consideration of disorder in treating temperature and annealing
effects.
\end{abstract}
\date{Version 1.9 -- \today}
\maketitle

\section{INTRODUCTION}
\label{intro}

The exciting possibilities of manipulating both the spin and the
charge of the carriers in semiconductors, in such a way that new
devices may be designed, have brought a lot of attention to the
study of diluted magnetic semiconductors (DMS) in the past ten
years or so. Even though the DMS have been known for a long time,
\cite{Furdyna88} it was the discovery of ferromagnetism in p-type
(In,Mn)As systems \cite{Ohno92} that spurred the research in this
field. This was even more so after the successful growth of
ferromagnetic (Ga,Mn)As alloys. \cite{Ohno96} This latter system
has become almost a paradigm in the field of DMS materials. It has
long been known that isolated Mn$_{\rm Ga}$ substitutional
impurities give rise to acceptor states around 0.1 eV above the
top of the valence band. Thus, the Mn atoms have a double
functionality in the Ga$_{1-x}$Mn$_x$As alloys: they provide both
(i) the magnetic moments, and (ii) holes to intermediate the
interaction between them. This somewhat simplistic view is much
more complex than it seems at first sight. Ferromagnetism in
Ga$_{1-x}$Mn$_x$As only occurs for large Mn concentrations of a
few percent. As a consequence, the acceptor levels form a band
which, due to the rather localized character of the defect state,
has a dispersion which is far from what would result from a free
quasi-particle picture. Moreover, the intrinsic disorder coupled
to this somewhat narrow band indicates that any theoretical
description based on an effective mass description should be
viewed with caution. To further complicate the issue, in order to
obtain the necessary high Mn concentrations the growth
temperatures cannot be too high, which causes a lot of defects to
be present in the samples, like Mn interstitials (Mn$_I$) and
arsenic antisites (As$_{\rm Ga}$). As a result, the critical
temperature and hole concentration, as a function of Mn
composition, are crucially dependent on the details of growth
conditions.
\cite{vanEsch97,Matsukura98,Ohno99a,Potashnik01,Edmonds02,Seong02,Asklund02,Yu02a,Yu02b,Potashnik02,Moriya03,
dosSantos02,dosSantos03}

In view of all these facts, it would be important to have a way to
estimate the Mn-Mn exchange interactions (i) with as few
assumptions as possible, (ii) which would treat the host and the
Mn impurities at the same level of accuracy, and (iii) which would
furthermore include the effects of disorder. This approach of
implicitly tracing out the holes degrees of freedom has been
implemented in a variety of ways based on self-consistent methods.
Van Schilfgaarde and Mryasov \cite{Schilfgaarde01} have performed
calculations of total energies, within the atomic spheres
approximation, to extract exchange couplings, $J$'s, for specific
(\emph{i.e.,} not randomly chosen) clusters of closely spaced Mn
ions; their results suggest a tendency of a decrease in $|J|$ when
more Mn atoms are added to nearby sites. More recently, Xu
\emph{et al.}\cite{Xu05} used muffin-tin orbitals to investigate
the dependence of the exchange coupling with the Mn-Mn distance at
much larger (8.3\%) concentrations of Mn atoms; they found a
considerable scatter in the values of the exchange couplings. 
% Changes here
In a series of theoretical studies, in which the effect of
randomness/disorder is described by the coherent-potential
approximation (CPA), Kudrnovsk\'y \emph{et al.}
\cite{Kudrnovsky04a,Kudrnovsky04b} and Bergqvist \emph{et al.}
\cite{Bergqvist04} have used a tight-binding linear muffin-tin
orbital method, together with the magnetic force theorem, to study
the dependence of the Mn-Mn exchange couplings and critical
temperatures with the concentration of Mn impurities in III-V and
group IV DMS. Also, Sato \emph{et al.} \cite{Sato03,Sato04} have
used muffin-tin type potentials together 
with a KKR-CPA approach to study Curie temperatures and exchange interactions in
III-V DMS. Moreover, Sandratskii and Bruno
\cite{Sandratskii02,Sandratskii03,Sandratskii04} have used the
augmented-spherical-wave method within the local-density
approximation to investigate exchange interactions, Curie
temperatures and the influence of the clustering of Mn impurities
in (Ga,Mn)As. One should notice that 
%that the lack of diretionality renders 
the use of non-full potential muffin-tin--style
approaches is not adequate to treat the
electronic structure of covalent semiconductor systems such as
(Ga,Mn)As DMS. Furthermore, we will show that disorder plays an
important role which may not be adequately treated by simple
effective-medium approaches such as the virtual crystal
approximation (VCA) or CPA.

In this work we perform large supercell total energy calculations,
based on {\it ab initio} density functional theory (DFT) methods.
Within this approach, we treat disorder configurations in which
the Mn atoms randomly replace Ga atoms. By considering two, three,
and four Mn atoms in a supercell with 128 atoms, we cover three Mn
concentrations, 3.1\%, 4.7\%, and 6.3\%, and present results for
the effective exchange interactions, $\JMn$, between two Mn atoms
which are $n$-th neighbors in the Ga sublattice, with $1 \le n \le
6$. Also, in a few cases the Mn atoms are placed in predetermined
positions, in order to compare the exchange coupling of two
nearest-neighbor Mn atoms in the presence of other Mn atoms,
placed at various separations. The present results indicate a
clear decrease in the magnitudes of the $\JMn$ with the Mn
concentration $x$; from now on, the Mn-Mn superscript in $\JMn$
will be omitted, in order to simplify the notation.

The work is organized as follows. In Section II we provide a brief
description of the calculational procedure used in the present
study. Results and discussion are left for Section III, and Section IV
summarizes our findings and conclusions.

%%%%%%%%%%%%%%%%%%%%%%%%%%%%%%%%% Section II %%%%%%%%%%%%%%%%%%%%%%%%%%%%%

\section{Calculational method}
\label{calcs}

We have performed total energy calculations based on the
density-functional theory (DFT) within the generalized-gradient
approximation (GGA) for the exchange-correlation potential, with
the electron-ion interactions described using ultrasoft
pseudopotentials.\cite{Vanderbilt90} A plane wave expansion up to
230 eV as implemented in the VASP code \cite{Kresse93} was used,
together with a 128-atom fcc supercell and 4 $L$-points for the
Brillouin zone sampling; these $L$-points are non-equivalent, due
to the presence of Mn impurities. The positions of all host GaAs atoms as well as substitutional Mn in
the supercell were relaxed until all the forces components were
smaller than 0.02 eV/\AA; our GGA lattice parameter for undoped 
GaAs turned out to be 5.74 \AA, which is in accordance with other estimates, e.g., that of Ref. \onlinecite{Mahadevan04a}. For a 128 atoms supercell, we consider a
variety of configurations with 2, 3 and 4 Mn atoms, corresponding
to concentrations of 3.1\%, 4.7\%, and 6.3\%, respectively. Since
calculations for \emph{all} possible disorder configurations with
more than 2 Mn atoms per cell is prohibitively costly in terms of
computer time, we have considered \emph{typical} configurations,
as generated through the Special Quasi-random Structures (SQS)
algorithm.\cite{Wei90} A configuration $\sigma$ is generated by
placing the Mn atoms at Ga sublattice sites (64 possible sites).
We then calculate the pair correlation functions, up to the
sixth-neighbor, given by:
\begin{equation}
\Pi_m(\sigma) = \frac{1}{64 Z_m} \sum_{i,j} \Delta_m(i,j)S_iS_j.
\end{equation}
\noindent Here $\Pi_m$ is the $m$th-neighbor pair correlation
function, $Z_m$ is the number of $m$th-order neighbors to a site,
$\Delta_m(i,j)$ is 1 if sites $i$ and $j$ are $m$th-order
neighbors, and zero otherwise; and $S_i$ is a variable taking
values 0, if site $i$ is occupied by Ga, and 1 if it is occupied
by Mn. For a perfectly random ($R$) distribution of Mn atoms, the
pair correlation function does not depend on $m$, $\Pi_m(R) =
x^2$, where $x$ is the Mn concentration. For a given configuration
we calculate the deviation from randomness as

\begin{equation}
\delta\Pi(\sigma) = \sum_m (\Pi_m(\sigma) - \Pi_m(R))^2.
\end{equation}

The above quantity indicates how random the $\sigma$ configuration
is. We perform an exhaustive search over all possible
configurations and choose to work with the ones with lowest
$\delta\Pi$.

For each chosen disorder configuration, we adopt the following
strategy within our DFT-GGA calculations. As an initial guess, we
take all valence electrons of each Mn atom aligned with each
other, corresponding to $S=5/2$ as expected in a $d^5$
configuration, and calculate the total energies for this
configuration, as well as for an increasing number of
flipped Mn total spins. The energy differences with respect to the
aligned states, $\{\Delta E\}$, are then described by an effective
Heisenberg model with appropriate first-, second-, and so forth,
up to sixth-nearest-neighbor Mn-Mn couplings, $J_n,\ n=1-6$; see
Sec.\ \ref{results} and Appendix for details. This procedure has
been applied \cite{daSilva03,daSilva04} to the case of two Mn
atoms in a supercell with 128 sites, and we were able to infer the
dependence of the effective couplings with both the Mn-Mn distance
and direction. 
We have found \cite{daSilva03,daSilva04} that the calculated $J_n$ 
exchange couplings lead to a Mn ferromagnetic state, with the holes 
forming a relatively dispersionless impurity band, and therefore that a 
conventional free-electron--like RKKY interaction should be ruled 
out as the origin of the Mn-Mn ferromagnetic coupling. 

One should notice that for two Mn atoms, if the spins are treated
quantum mechanically the above mentioned energy difference
corresponds to that between the state with total spin 5/2 and the
singlet one, leading to $J_n^{\rm(Q)}=\Delta E_n/15$.
If the spins are treated classically, $J_n^{\rm(Cl)}=2\Delta E_n/25$
and the two approaches are entirely equivalent, apart from an
overall multiplicative factor of 1.2. For more than two Mn atoms,
we consider classical spins and note that this approximation,
though not capturing full details of the excitation spectra, is
still able to provide overall trends of the low energy magnetic
excitations for a finite number of spins.

As a final methodological comment, we note that we have checked
for spin-orbit effects (in the case of two Mn atoms) through the
projector augmented-wave (PAW) method \cite{Kresse99}, and found a
change from 0.29 eV to 0.24 eV in the total energy difference
between the excited antiferromagnetic and ground state
ferromagnetic Mn-spin alignements. Although a systematic study in
this sense would certainly be important, this is beyond the scope
of the present study and we have chosen to ignore spin-orbit
effects in the total energy calculations presented in this work.
We believe this approximation would not alter the general
conclusions of the present study. Moreover, other possibilities, 
such as a non-collinear ferromagnetism,\cite{Schliemann02,Zarand02} 
have not been considered at this stage.

%%%%%%%%%%%%%%%%%%%%%%%%%%%%%%%%% Section III %%%%%%%%%%%%%%%%%%%%%%%%%%%%%

\section{Results and Discussion}
\label{results}

\subsection{Two Mn atoms}

Let us first discuss the case of two Mn substitutional atoms in the 128-site supercell.
We considered \emph{all} configurations corresponding to all inequivalent positions within the supercell, i.e., Mn-Mn
distances varying from 4.06 \AA\ up to 11.48 \AA. Our total energy results yield a Mn-Mn ferromagnetic ground state in all cases.
The relevant Heisenberg Hamiltonian in this case is
\begin{equation}
H=J_n {\bf S}_{\bf i} \cdot {\bf S}_{\bf i + n},
\label{H2}
\end{equation}
for each relative position in the supercell, where, for the sake of comparison with the cases of three and four
 Mn atoms (see below), {\bf S} is taken as a classical spin of magnitude 5/2; {\bf n} is a vector connecting
 $n$th--nearest-neighbor Mn atoms replacing Ga atoms.
The estimates for $J_n$ thus obtained are displayed in the second column of Table \ref{table1}.

\begin{table}
\caption{\label{table1}Estimates for the effective exchange
coupling, $J_n$, in meV, between $n$th--nearest-neighbor Mn spins,
${\bf S}_{i}$ and ${\bf S}_{j}$, for different Mn concentrations,
$x$. In the case of 2 Mn atoms ($x=3.1\%$), $J_n$ is unique for a
given $n$. For 3 Mn atoms ($x=4.7\%$) in a supercell with 128
sites, we have performed calculations for 10 different SQS
disorder configurations, and $J_n$ is given by the average over
the configurations in which two Mn sites are $n$th-neighbors; the
number of such configurations are shown in square brackets, and
the error bars are calculated as standard deviation of averages.
In the case of 4 Mn atoms ($x=6.3\%$), we show the results for two
configurations (see text); note that sometimes a specific
configuration would not accommodate the pertinent $J_n$.}
\begin{ruledtabular}
\begin{tabular}{ccccc}
     $n$ & $x=3.1\%$ & $x=4.7\%$ & $x=6.3\%$ & $x=6.3\%$\\
\hline\\
     1   & $-23.2$& $-18.2\pm 1.5 $ [7] & $ -12.6 $ & $ -13.0 $\\
     2   & $-10.4$& $-3.8 \pm 1.8 $\ [4]&      -     & $ -4.7  $\\
     3   & $-13.6$& $-6.6 \pm 2.7 $\ [7]& $ -2.8  $ & $ -6.0  $\\
     4   & $-5.6 $& $-3.6 \pm 0.8 $\ [4]& $ -4.8  $ &     -     \\
     5   & $-2.6 $& $+0.4 \pm 0.7 $\ [5]& $ +0.1  $ & $ -1.3  $ \\
     6   & $-4.4 $& $-1.9 \pm 0.7 $\ [2]&      -    &     -    \\
\end{tabular}
\end{ruledtabular}
\end{table}

As previously noted,\cite{daSilva03,daSilva04} the resulting
Mn-Mn ferromagnetic effective coupling in Ga$_{1-x}$Mn$_x$As is
essentially intermediated by the antiferromagnetic coupling of
each Mn spin to the quasi-localized holes. Also, the observed
non-monotonic behavior of $J_n$ should be attributed to the
anisotropic character of the effective interaction. Moreover,
$|J_n|$ essentially decreases \cite{daSilva03,daSilva04} with Mn-Mn
separation and vanishes above $\sim 11.5\, \text{\AA}$.

\subsection{Three Mn atoms}

\begin{figure}[h]
{\centering\resizebox*{8.0cm}{!}{\includegraphics*{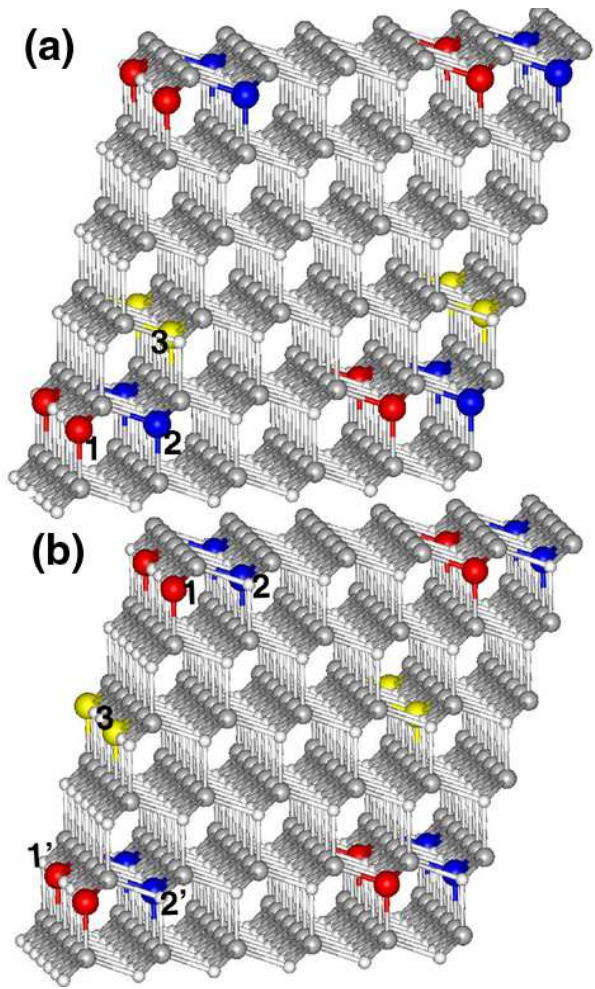}}}
\caption{(Color online) A pictorial view of two possible realizations of disorder
for three Mn atoms in a 128-site supercell ($x=4.7\%$). Ga sites are represented by the smaller spheres,
As sites by the middle-sized ones, and Mn atoms by the largest ones. For clarity, supercells are repeated along
the different cartesian directions. The three nonequivalent Mn atoms are shown as different shades of gray
(blue, red, and yellow in the color version).}
\label{3Mn}
\end{figure}

In the case of three Mn atoms in a supercell with 128 sites, we
have performed calculations for 10 different disorder
configurations. Figure \ref{3Mn} shows two SQS illustrative
configurations: in (a) the 3 Mn atoms are somewhat clustered
together, whereas in (b) two are nearest neighbors and the third
is farther apart.

\begin{figure}[h]
{\centering\resizebox*{8.0cm}{!}{\includegraphics*{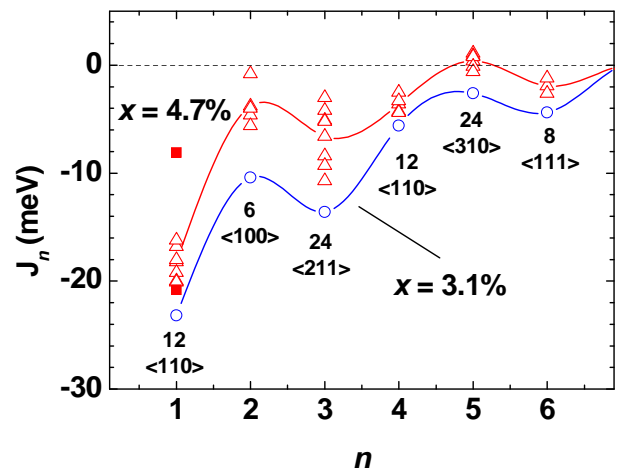}}}
\caption{(Color online) The $n$th--nearest-neighbor exchange
coupling as a function of $n$ for $x=3.1\%$ and $x=4.7\%$ (with
data displayed for 10 SQS configurations, see text). Full curves
are guides to the eye (for $x=4.7\%$ the full line goes through
average values of $J_n$). Filled squares for $J_1$ correspond to
extreme values obtained for the non-SQS configurations; see text.
Also shown is the multiplicity of each $n$-th neighbor pair in a given direction $<hkl>$.}
\label{3Mn-Jn-vs-n}
\end{figure}

For each disorder configuration, the relevant Heisenberg Hamiltonian
must contemplate the possibility of interactions occurring not only
amongst spins within the supercell, but between one spin in the
supercell and the different images in neighboring supercells
(periodic boundary conditions effects, PBCE's).
In actual fact, depending on the disorder configuration, the same
pair of spins may be $j$-th nearest neighbors within the supercell and
$k$-th nearest neighbors when the images are considered.
One can therefore write the Hamiltonian as
\begin{equation}
H=\sum_n \sum_{i<j} w_{ij}(n) J_n\ {\bf S}_i\cdot {\bf S}_j,
\label{H3}
\end{equation}
where the $w_{ij}(n)$ are geometrical weights taking into account
PBCE's. For a given configuration, one expects most of the $w$'s
to vanish; also, we set $w=0$ if the distance between the Mn atoms
is larger than 11.5 \AA, as previously established.\cite{daSilva04}
In the Appendix we discuss the Hamiltonian for the two SQS
configurations of Fig.\ \ref{3Mn}.
We then calculate the total energies for different Mn spin
configurations: with all spins aligned, with only one spin
reversed, either in site 1, 2, or 3, and so forth, increasing
the number of spin flips, until one has the same number of unknowns
($J_n$) as equations (namely, the corresponding energy differences
with respect to the aligned state).

\begin{figure}[ht]
{\centering\resizebox*{5.5cm}{!}{\includegraphics*{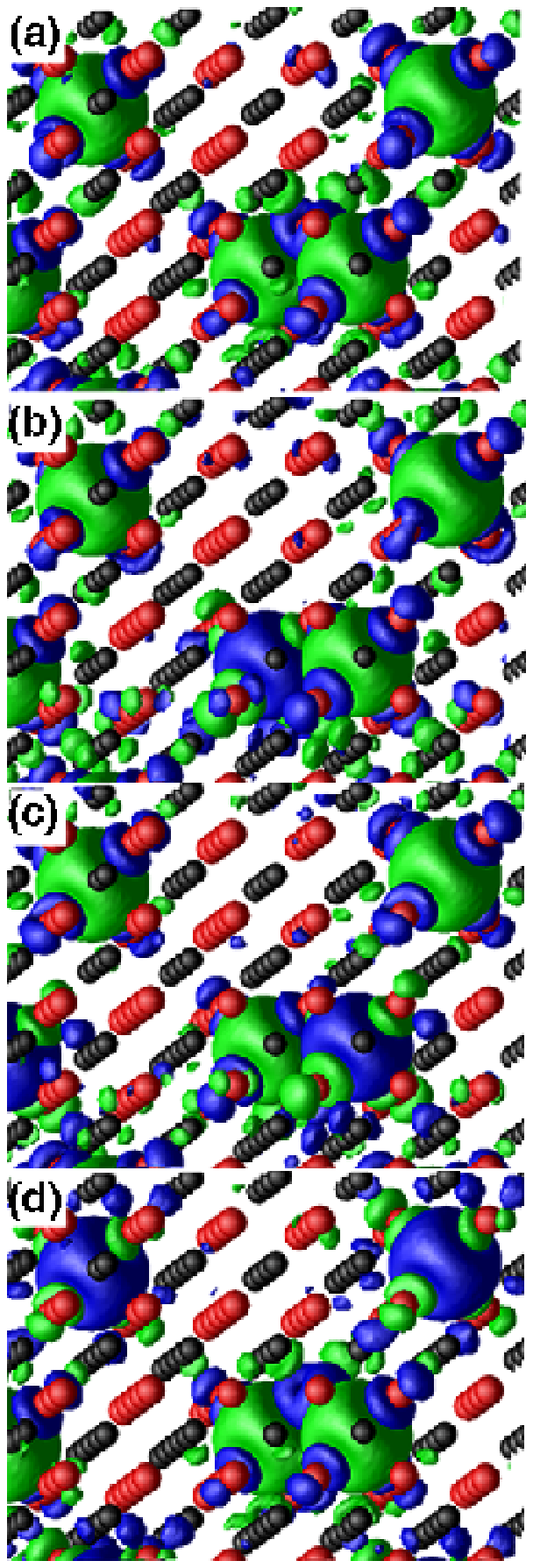}}}
\caption{(Color online) Isosurfaces for the net local
magnetization $m({\bf r})$ (see text for definition) in the case
of three \MnGa defects [for the configurations depicted in Fig.
\ref{3Mn}(a)], with (a) all spins aligned and (b)-(d) only one
flipped spin. The green surface corresponds to a value of $+ \ 0.005 \ e/\text{\AA}^3$, and the blue
surface to $- \ 0.005 \ e/\text{\AA}^3$, with $e$ being the electron charge. The black
(red) spheres denote the Ga (As) atoms.
} \label{3Mn-iso}
\end{figure}

It is instructive to lay out $J_n$ as a function of $n$ for the 10
SQS realizations of disorder (three Mn atoms), as shown in Fig.
\ref{3Mn-Jn-vs-n}; for comparison, we show the results for
$x=3.1\%$ in the same figure. One can see that the overall trend
of $J_n$ with $n$, observed in the case of two Mn spins, is
maintained in this case, with the non-monotonic behavior still
being due to effects of directionality, i.e., the exchange coupling depends 
not only on the distance between the pair of Mn atoms, but also on their relative 
direction with respect to the bonds of the host GaAs. 
Here we should mention that our 128-atom supercell total energy results for the ferro- 
and antiferromagnetic states are in overall agreement with the corresponding 64-atom supercell total energy results of Mahadevan \emph{et al.}.\cite{Mahadevan04b} 
The corresponding average values of $J_n$, for each $n$, are shown in
the third column of Table \ref{table1}. It is interesting to note
that all $J_n$ decrease (in absolute value) as the concentration
of Mn atoms increases from 3.1\% to 4.7\%. While at first sight
this may seem an unusual behavior, one should have in mind that
the effective Mn-Mn interaction is hole-mediated, thus sensitive
to the hole density.

In order to assess the effects of clustering in a systematic way,
we have also considered non-SQS configurations in which two Mn
atoms are first neighbors, and a third Mn atom is placed in
positions corresponding to fifth-, third-, and first-neighbor of
the pair: we found that $J_1=-20.8$ meV, $-17.3$ meV, and $-8.1$
meV, respectively; the extreme values are shown in Fig.\
\ref{3Mn-Jn-vs-n} as filled squares. Thus, \emph{clustering tends
to weaken the magnitude of the nearest-neighbor coupling}. One may
attribute this behavior as most likely resulting from the Coulomb
repulsion between the holes, which leads to their delocalization
as the Mn atoms group together, being therefore detrimental of
their role as mediators of ferromagnetism.

\begin{table}
\caption{\label{table2} Total energies from ferromagnetic SQS
configurations, labelled from $\ell = 1$ to 10, with respect to the total
energy of the configuration corresponding to three
nearest-neighbor Mn atoms clustered together. The effective Heisenberg Hamiltonian can be written in the form $H=\alpha J_i {\bf S}_1\cdot{\bf S}_2+ \beta J_j {\bf S}_1\cdot{\bf S}_3 + \gamma J_k {\bf S}_2\cdot{\bf S}_3$, such that the entries in the third column are $\{i^\alpha j^\beta k^\gamma \}$}
\begin{ruledtabular}
\begin{tabular}{cccc}
       $\ell$-th state & $E_\ell$ (eV) & $\{i^\alpha j^\beta k^\gamma \}$ \\
\hline\\
                  1 & 0.054 & $\{1^1 3^1 4^2\}$    \\
                  2 & 0.077 & $\{1^1 4^2 5^2\}$    \\
                  3 & 0.085 & $\{1^1 3^1 3^1\}$    \\
                  4 & 0.090 & $\{1^1 3^1 6^4\}$    \\
                  5 & 0.091 & $\{1^1 3^1 5^2\}$    \\
                  6 & 0.103 & $\{1^1 2^1 3^1\}$    \\
                  7 & 0.125 & $\{1^1 2^1 5^2\}$   \\
                  8 & 0.158 & $\{2^1 4^2 6^4\}$   \\
                  9 & 0.176 & $\{3^1 4^2 5^2\}$    \\
                 10 & 0.227 & $\{2^1 3^1 5^2\}$    \\
\end{tabular}
\end{ruledtabular}
\end{table}

If, on the one hand, clustering tends to decrease the magnitude of
the nearest-neighbor exchange, on the other hand it leads to the
energetically most stable configuration; this is in agreement with 
recent results from calculations restricted to pairs of transition 
metals.\cite{Mahadevan05a} In Table \ref{table2} we
display the energies of calculated ferromagnetic SQS
configurations relative to the clustered one in which the three Mn
atoms are first-nearest neighbors. We note that the SQS
configurations labelled from 8 to 10, which have the highest total
energies of the set, correspond to cases in which there are no
first-neighbor pairs of Mn atoms.
Since Ga$_{0.97}$Mn$_{0.03}$As is only stable at growth temperatures
in the range 200--300\,C,\cite{Ohno99a,MacDonald05} the scale of
energies shown in Table \ref{table2} indicates that not many
configurations can be thermally activated. Clearly, there are several
other mechanisms at play -- such as mobility of Mn atoms, possibility
of trapping on interstitials, and so forth --, which are not included in the
present approach, and will determine the final distribution of Mn atoms.

Figures \ref{3Mn-iso}(a)-(d) show the net magnetization $m({\bf
r}) \equiv \rho_{\up}({\bf r}) - \rho_{\dn}({\bf r})$, where
$\rho_\sigma$ is the total charge density in the
$\sigma$-polarized channel, for three Mn atoms with all spins
aligned and for only one flipped spin, for the configuration
depicted in Fig. \ref{3Mn} (a). Note that the densities on the
upper right and upper left corners in each figure are related to
a Mn atom and its image in a neighboring supercell. Similarly to the $m({\bf r})$ of one\cite{daSilva03} and two
Mn impurities\cite{daSilva04} in a supercell, near each
Mn atom the local magnetization has a  $d_{\sigma}-$like character, whereas close to
the As neighbors, the character changes to $p_{\bar{\sigma}}-$like, where $\sigma=(\uparrow or \downarrow$)
and $\bar{\sigma}=(\downarrow or \uparrow)$. Also, $m({\bf r})$ has a rather localized
character. The flipping of spins introduce nodes on the $m({\bf r})$ and subtle changes
mostly on the orientation of the $p-$like lobes. Close to the Mn atoms, however,
the local magnetization is not very sensitive to the flips.

\subsection{Four Mn atoms}

For four Mn atoms, we have considered only two disorder
configurations, chosen according to the SQS algorithm. A
Hamiltonian similar to Eq.\ (\ref{H3}) may be written, with the
addition of terms involving the fourth spin, having in mind that
the interactions with spins on image sites are more frequent in
this case.

\begin{figure}[ht]
{\centering\resizebox*{6.6cm}{!}{\includegraphics*{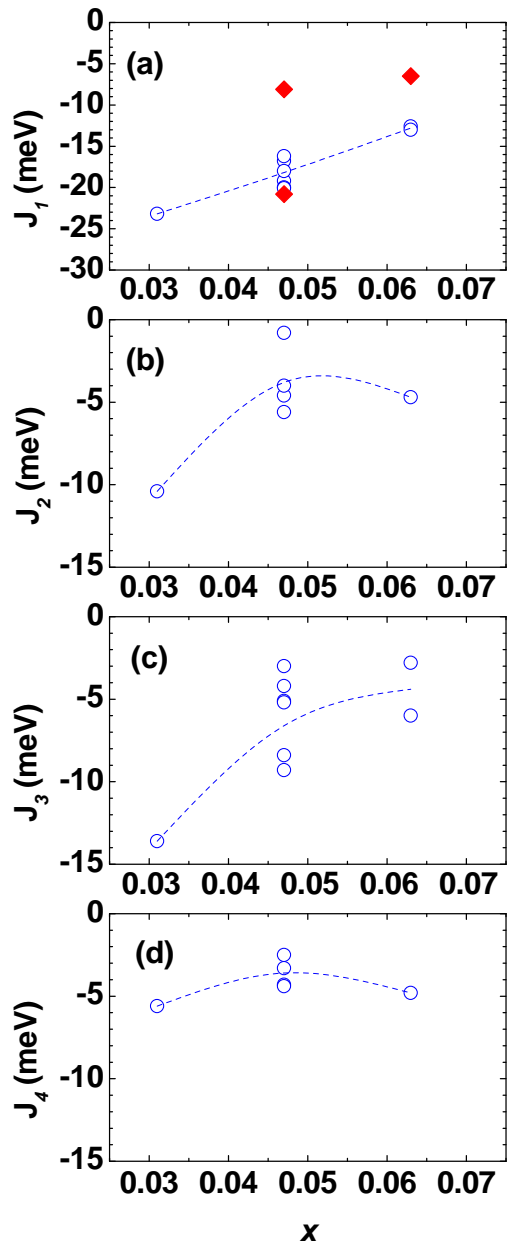}}}
\caption{(Color online)Dependence of: (a) $J_1$, (b) $J_2$, (c)
$J_3$, and (d) $J_4$ with the concentration of Mn atoms. For
$x=3.1\%$, $J_n$ is unique for a given $n$. Values for the SQS
configurations are shown as empty circles, while the filled
diamonds correspond to the extreme values obtained for the non-SQS
configurations; see text. Dotted curves are guides to the eye
through the average values of $J_n$.} \label{J-vs-x}
\end{figure}

For instance, in one of the calculated SQS configurations, the
effective Hamiltonian becomes
\begin{eqnarray}
H=&2J_5 \sv_1\cdot\sv_2+J_1 \sv_1\cdot\sv_3+ J_3 \sv_1\cdot\sv_4\nonumber\\
&2J_4 \sv_2\cdot\sv_3+ J_3 \sv_2\cdot\sv_4+ J_1 \sv_3\cdot\sv_4,
\label{H4a}
\end{eqnarray}
where the absence of a $J_2$ second-neighbor interaction should be
noticed.  Calculations of total energies for all Mn spins
parallel, and for the four possible single flips, lead to four
excitation energies, from which the $J_n$'s ($n\neq 2$) may be
inferred. Analogous considerations apply to the other SQS
configuration. The results are shown in columns 4 and 5 of Table
\ref{table1}. One sees that the overall tendency of $J_n$ is to
decrease in magnitude as $n$ is increased, in a pattern similar to
that for smaller concentrations, though the dispersion cannot be
properly assessed due to the scarcity of data. We also note that,
as in the case of three Mn atoms, calculations with a non-SQS
configuration with the four Mn atoms clustered together indicate
that clustering decreases the magnitude of the first-neighbor
$J_1$ exchange coupling: $J_1 = - 6.5$ meV in this case, which
should be compared with the $- 12.6$ meV and $- 13.0$ meV values
of Table \ref{table1}.

\subsection{The dependence of $J_n$ with the concentration}

The data in Table \ref{table1} can also be used to discuss the
dependence of $J_n$ with $x$, for a given $n$. In Fig.\
\ref{J-vs-x} we plot $J_1$, $J_2$, $J_3$ and $J_4$ as functions of
$x$. For the case of $J_1$, we also show (as filled symbols)
three values obtained for the non-SQS configurations: two as
mentioned before, in the case of three Mn atoms, and the one
corresponding to four Mn atoms clustered together as first
nearest-neighbors.

From Fig.\ \ref{J-vs-x}, we see that, in most cases, the
magnitudes of the exchange couplings decrease as the concentration
of Mn atoms is increased. Further, this decrease may be quite
significant; for instance, the magnitude of the average $J_1$
decreases by the order of 50\% when one roughly doubles the
concentration from 3.1\%. We also see that for the configurations
in which the Mn atoms are clustered together, $|J_1|$ also
decreases as $x$ is increased. This overall decrease with $x$ can
be taken as numerical evidence that a steady increase in the
concentration of Mn atoms is not sufficient to rise the critical
temperature, since the exchange couplings will eventually be
weakened. Clearly other effects may be playing important roles.
For instance, within our present approach, the hole density is
assumed to be the same as that of Mn atoms, which, as mentioned in
the Introduction is not really the case. The presence of Mn
interstitials and Mn-As complexes also need to be taken into
account in order to reach a quantitative agreement. Nonetheless,
one expects that the trends unveiled here are indicative of the
actual experimental situation. 

It is important to have in mind that several theoretical works 
have previously examined the dependence of the exchange couplings
with the Mn-Mn separation or with the Mn concentration.\cite{Kudrnovsky04a,Kudrnovsky04b,Bergqvist04,Sato03,Sato04,Sandratskii02,Sandratskii03,Sandratskii04,Xu05} Some predict an alternating sign for the exchange coupling, but these predictions should be taken with extreme care, since these theoretical calculations are based on non--full-potential 
muffin-tin--type potentials which are not reliable to treat the electronic structure of covalent semiconductor systems such as (Ga,Mn)As DMS. Also, disorder quite certainly is not adequately
taken into account within simple effective-medium approaches such
as VCA or CPA, as fluctuations in the Mn positions essentially
lead to variations in the Mn-Mn exchange-coupling parameters, as
apparent from Fig.\ \ref{J-vs-x}.

%%%%%%%%%%%%%%%%%%%%%%%%%%%%%%%%% Section IV %%%%%%%%%%%%%%%%%%%%%%%%%%%%%

\section{Conclusions}
\label{concs}
We have performed {\it ab initio} total energy density-functional calculations for
two, three, and four substitutional Mn atoms in a 128 atoms supercell, corresponding
to concentrations of 3.1\%, 4.7\%, and 6.3\%, respectively. In this way, we have treated
the host and the Mn impurities on equal footing. The effects of disorder have been assessed
at different levels of approximations, depending on the concentration of Mn atoms: for
$x=3.1\%$, \emph{all} possible non-equivalent positions of the Mn atoms have been considered;
for $x=4.7\%$, ten non-equivalent configurations have been generated through the SQS algorithm,
while three specific ones have also been considered in
order to discuss the effects of clustering; and, for $x=6.3\%$, two SQS and one non-SQS configurations have been
investigated. While the relation between the densities of holes and of Mn atoms is one of the yet unsolved
issues in the context of DMS, here we have assumed that each Mn atom provides one hole; since our results
relate to general trends, they may be carried over to the actual experimental situation of only a
fraction of Mn atoms contributing with holes.
It is also interesting to note that the cut-off of 11.5 \AA\ (which would correspond to $x\simeq 0.042$) imposed on the range of Mn-Mn exchange couplings would appear to be in direct contradiction with experimental data by Edmonds et al.\cite{Edmonds02}, according to which  ferromagnetism is seen for dopings as low as $\sim 0.015$ (where one would have essentially no compensation). Since the site percolation threshold\cite{Stauffer92} for FCC lattices is 0.20,  for the Ga FCC sublattice in (Ga,Mn)As, the concentration cut-off for ferromagnetic order would be of the order of $0.20 \times 0.042 = 0.0084$, i.e., $x\simeq 0.84\%$, indicating that there is no contradiction with the measurements of Edmonds et al.\cite{Edmonds02}

We have focused mainly on the effective exchange interaction
between Mn spins, by mapping the spectra of magnetic excitations
(spin flips) onto a classical Heisenberg Hamiltonian with coupling
constants $J_n$, ranging from first ($n=1$) to sixth ($n=6$)
nearest neighbors. The effects of clustering on the
nearest-neighbor pair-exchange coupling, $J_1$, have been
investigated by examining specific (i.e., non-random)
configurations with three and four Mn atoms in the 128-site
supercell: We have established that clustering tends to weaken the
magnitude of the nearest-neighbor exchange coupling.
On the other hand, we have
found that clustered structures of Mn atoms have the lowest total
energies, a result which may be of importance in a realistic
discussion of annealing and/or diffusion effects. From
calculations on random configurations we have also been able to
determine the behavior of $J_n$ with $x$, for fixed $n$: in most
cases the exchange couplings get weaker as the concentration of Mn
atoms is increased. 
This is consistent with the experimentally
observed fact that there is an optimum range of Mn concentrations
(whose quantitative determination requires a careful consideration
of other disorder effects) in which the critical temperatures are
the highest.

For fixed Mn density, we have found that the calculated $J_n$
favor a ferromagnetic ground state, and have decreasing magnitude
as the distance between spins increases (cf. Table I and Figs. 2
and 4). The non-monotonic behavior is attributed to directionality
effects; by the same token, deviations in the sign of $J_n$ were
found only at large $n$ (=5), when its magnitude is already
greatly reduced with respect to the nearest neighbor value. The
discrepancy of the present results with respect to recent
calculations by Xu \emph{et al.},\cite{Xu05} may be attributed to
the fact that their muffin-tin calculations are not full
potential; they therefore do not fully reproduce the crucial role
played by the directional $sp^3$ bonds and by the hole $p$-states.
Also, due to the quite significant variations of the calculated
exchange couplings with configurations and Mn concentration, we
emphasize that estimates of the critical temperature obtained via
exchange couplings thus obtained are clearly open to question. We
should also stress that the present results corroborate that the
Mn-Mn ferromagnetic effective coupling in Ga$_{1-x}$Mn$_x$As is
intermediated by \emph{localized} holes leading to an
antiferromagnetic (non-RKKY) coupling of each Mn spin, as
previously noted,\cite{daSilva03,daSilva04} and recently confirmed
experimentally.\cite{Sapega05} 
Therefore, the inescapable
conclusion is that the main feature of a conventional
free-electron--like or perturbative RKKY interaction should be
ruled out\cite{Priour04,Timm05} in the case of Ga$_{1-x}$Mn$_x$As.

As a final point, some comments regarding future perspectives are
in order. From one side, investigations using a similar procedure
as employed here (where the disorder is explicitly included) of
how impurities, such as interstitial Mn and As anti-sites, alter
the effective exchange interactions are relevant. For a given Mn
configuration, it should be interesting to see how the results
depend on the relative position of the defects. On the other hand,
our results raise some questions whose answers are not completely
trivial: (i) The fact that the effective exchange interactions
change with the Mn configuration make it clear that the use of a
Heisenberg model, at least a simple one where only
pair-interactions are considered, should be viewed with caution.
It is not obvious that extensions of the Heisenberg model to
triplets or even larger cluster interactions will remedy this
fact; (ii) The use of {\it ab initio} calculations has been very
important in order to provide a correct picture of the electronic
structure of these systems. One of its great merits is the
possibility of obtaining model-free results. However, whenever one
needs to make predictions about the critical temperature ($T_c$),
models have to be used. For instance, from {\it ab initio} results
one may extract effective exchange parameters, as in the present
work, and then via mean field or more sophisticated methods, like
Monte Carlo calculations, it is possible to calculate $T_c$. Two
crucial steps in this procedure are questionable. The first one is
the use of a Heisenberg model, as already mentioned. The other is
the use of a small supercell approximation. Calculating the
critical temperature via any effective methodology that is based
on {\it small} supercell {\it ab initio} calculations, even if
this effective approach allows the search of a large number of
distinct configurations, has a great risk of being nonsense,
since, as we have shown, the exchange interactions depend
sensitively on the Mn distribution. The root of the above problems
is the necessity of introducing a model hamiltonian in order to
extract excited states of the system associated with spin
excitations. A possible solution to this problem could be the use
of a semi-empirical hamiltonian with a tight-binding descrition
for the host material coupled with a many-body, atomic-like
description for the Mn atoms. The manyfold of low-energy states
representing the different Mn spin orientations, that will be
obtained upon diagonalization of such a hamiltonian\cite{daSilva95},
will replace the states obtained via the effective (but
questionable) Heisenberg hamiltonian.

\begin{acknowledgments}
Partial financial support by the Brazilian Agencies CNPq, CENAPAD-Campinas, Rede Nacional de
Materiais Nanoestruturados/CNPq, FAPESP, FAPERJ, and Millenium Institute for Nanosciences/MCT
is gratefully acknowledged.
\end{acknowledgments}

\appendix*
\section{}

Here we discuss the case of three Mn atoms, for the two disorder
configurations displayed in Figs. 1(a) and 1(b), and chosen
according to the SQS algorithm. For the configuration in Fig.
1(a) ($\ell=4$ in Table II), the Hamiltonian [see Eq.\ (\ref{H3})] may be written, having
in mind that the interactions with spins on image sites are to be
taken into account, as
\begin{eqnarray}
H = J_1 \sv_1\cdot\sv_2 + J_3 \sv_2\cdot\sv_3+ 4J_6
\sv_1\cdot\sv_3
\end{eqnarray}
where the absence of second-neighbor, fourth-neighbor, and
fifth-neighbor interactions should be noticed.  

In a similar way, for the configuration in Fig. 1(b) ($\ell=9$ in Table II), the
Hamiltonian is given by
\begin{eqnarray}
H =  J_3 \sv_1\cdot\sv_2 +2J_4 \sv_1\cdot\sv_3 + 2J_5
\sv_2\cdot\sv_3
\end{eqnarray}
where one notes the absence of first-neighbor, second-neighbor,
and sixth-neighbor interactions.

One may perform DFT-GGA calculations, and obtain the total
energies for SQS configurations with all Mn $S=5/2$ atoms aligned
with each other, as well as for an increasing number of flipped Mn
total spins. The total-energy differences with respect to the
aligned states, $\{\Delta E\}$, may then be obtained via an
effective classical Heisenberg model with appropriate $J_n$
exchange couplings up to $n = 6$.

For the disorder configuration in Fig. 1(a), noticing that
classically one has $\sv_i\cdot\sv_j = \pm\frac{25}{4}$, it is
straightforward to obtain, using eq. (A.1), for the total energies
of configurations with appropriate flipping of Mn total spins
\begin{eqnarray}
E_0   &= \frac{25}{4} (+ J_1 + J_3 + 4J_6)\\
E_1^1 &= \frac{25}{4} (- J_1 + J_3 - 4J_6)\\
E_1^2 &= \frac{25}{4} (- J_1 - J_3 + 4J_6)\\
E_1^3 &= \frac{25}{4} (+ J_1 - J_3 - 4J_6),
\end{eqnarray}
where the lower index indicates the number of flipped spins (from
+5/2 to - 5/2), and the upper index labels which spin was flipped.
The differences in corresponding Heisenberg energies are therefore
\begin{eqnarray}
\Delta_{1-0} &= - \frac{25}{4} (2J_1 + 8J_6)\\
\Delta_{2-0} &= - \frac{25}{4} (2J_1 + 2J_3)\\
\Delta_{3-0} &= - \frac{25}{4} (2J_3 + 8J_6),
\end{eqnarray}
and one may thus obtain $J_1 = - 16.2 \,$ meV, $J_3 = -3.0 \,$
meV, and $J_6 = - 1.2 \,$ meV from the calculated first principles
differences in total energies, i.e., $\Delta_{1-0} = 264 \,$ meV,
$\Delta_{2-0} = 241 \,$ meV, and $\Delta_{3-0} = 99 \,$ meV.

%\pagebreak

\bibliography{biblio-dms}

\end{document}